%% file: main.tex
\documentclass[letterpaper]{article}
\usepackage[preprint]{aaai2027}
\usepackage[hyphens]{url}
\usepackage{graphicx}
\usepackage{natbib}
\usepackage{caption}
\usepackage{booktabs}
\usepackage[table]{xcolor}
\usepackage{amsmath}
\usepackage{amssymb}
\DeclareUnicodeCharacter{00D7}{\ensuremath{\times}}
\DeclareUnicodeCharacter{2014}{---}
\DeclareUnicodeCharacter{2081}{\textsubscript{1}}
\DeclareUnicodeCharacter{2082}{\textsubscript{2}}
\DeclareUnicodeCharacter{2083}{\textsubscript{3}}
\DeclareUnicodeCharacter{2084}{\textsubscript{4}}
\DeclareUnicodeCharacter{2085}{\textsubscript{5}}
\DeclareUnicodeCharacter{2087}{\textsubscript{7}}
\DeclareUnicodeCharacter{2098}{\textsubscript{m}}
\DeclareUnicodeCharacter{2212}{\ensuremath{-}}
\DeclareUnicodeCharacter{222A}{\ensuremath{\cup}}
\DeclareUnicodeCharacter{22EE}{\ensuremath{\vdots}}
\DeclareUnicodeCharacter{22EF}{\ensuremath{\cdots}}
\DeclareUnicodeCharacter{2C7C}{\textsubscript{j}}
\DeclareUnicodeCharacter{03A9}{\ensuremath{\Omega}}

\title{Making Single-Cell Data Distillation Auditable: Traceable Real-Cell Coresets via Discrete Min--Max Selection}
\author{
Yaodi Luo\textsuperscript{\rm 2},\quad
Peize He\textsuperscript{\rm 2},\quad
Lingbei Meng\textsuperscript{\rm 1,2},\quad
Bowen Han\textsuperscript{\rm 2},\\
Zheng Lu\textsuperscript{\rm 2},\quad
Jianqing Zhu\textsuperscript{\rm 2}\corresponding,\quad
Lian Zhang\textsuperscript{\rm 3}\corresponding
}
\affiliations{
\textsuperscript{\rm 1}The Chinese University of Hong Kong, Shenzhen\\
\textsuperscript{\rm 2}Shenzhen Loop Area Institute, Shenzhen\\
\textsuperscript{\rm 3}Shenzhen Research Institute of Big Data, Shenzhen, China\\
Correspondence to: Lian Zhang (\texttt{zhanglian@sribd.cn}) and
Jianqing Zhu (\texttt{jianqingzhu@slai.edu.cn})
}

\begin{document}
\maketitle
\input{sections/abstract}

\input{sections/introduction}
\input{sections/related_work}
\input{sections/method}
\input{sections/experiments}
\input{sections/conclusion}

\clearpage
\bibliography{references}
\end{document}

%% file: sections/abstract.tex
\begin{abstract}
Large single-cell datasets are expensive to store, curate, and repeatedly
reuse for model training. Data distillation can reduce this burden by
building smaller training sets. However, many existing methods rely on
synthetic cells. These synthetic cells do not retain direct correspondence
with assayed cells and genes. This limits source-level inspection and
biological traceability. Moreover, real-cell expression matrices are often
sparse and noisy. In light of these challenges, we propose
\textbf{Minmax-CF}, a label-aware characteristic-function selector for
traceable single-cell data distillation. Minmax-CF formulates compression as
a discrete min--max selection problem over characteristic-function
directions. It uses entropy-regularized maximization to emphasize the least
preserved directions. Greedy minimization ranks cells and genes by how much
they reduce the resulting weighted error. The method alternates cell and
gene selection under explicit axis-specific budgets. Across five coarse-lineage benchmarks and five compression budgets, Minmax-CF retains 95.3\% of the Full-reference 
macro-F1 on average, with gaps that exceed one per-seed standard deviation. It also retains exact source-cell
indices and original gene symbols. Compared with size-matched synthetic
PCA-Centroid and Distribution Matching (DM) baselines, Minmax-CF achieves higher
coarse-lineage macro-F1 in 24 of 25 comparisons against each baseline. It
exceeds their average performance by 10.4\% and 17.4\%, respectively.
Retained cells can also be projected onto independently computed embeddings
for direct biological interpretation.
\end{abstract}

%% file: sections/introduction.tex
\section{Introduction}
\begin{figure}[t]
\centering
\includegraphics[width=\columnwidth]{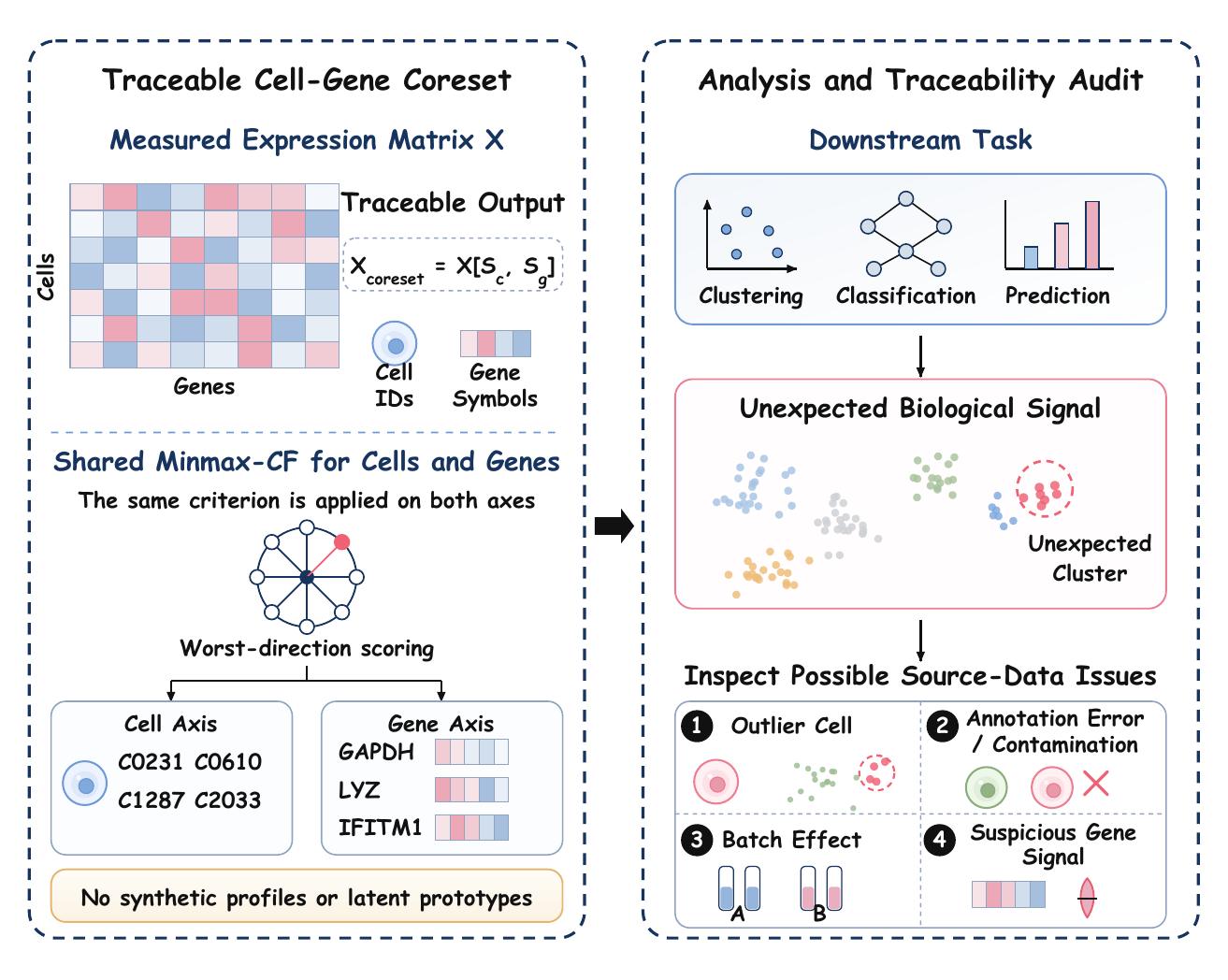}
\caption{\textbf{Source inspection of distilled data.} Minmax-CF preserves
measured cells and original genes. Their indices allow lookup in the source
matrix and metadata, but do not attribute predictions to retained cells.}
\label{fig:intro-concept}
\end{figure}
Single-cell datasets continue to grow in cell number and gene dimension. The
resulting matrices are costly to store and reuse. Model comparison,
hyperparameter search, annotation updates, and validation revisit the same
cells and genes. Principal component analysis provides a linear
representation. t-SNE and UMAP provide nonlinear embeddings
\citep{vandermaaten2008tsne,mcinnes2018umap}. Sketching and metacells
summarize cellular state space \citep{hie2019geometric,persad2023seacells}.
These methods support visualization and coverage. They do not by themselves
return a compact training set whose rows and columns remain tied to assayed
cells and measured genes. To overcome this gap, we need traceable single-cell data distillation, whose goal is to construct a small and reusable cell and gene coreset. Its entries should remain easy to inspect.

Source identity supports later inspection of retained training examples.
Synthetic profiles and centroids may provide compact supervision. They do
not identify one assayed source cell. A coreset of measured cells preserves
the sample, annotation, quality control, sequencing record, and count
profile. These records support follow up when a rare population,
perturbation response, or marker signal behaves unexpectedly. The retained
training data can therefore be inspected, providing an evidence basis for
further attribution, error analysis, and biological interpretation.

Making the distilled object traceable changes the optimization problem.
Continuous synthesis can move in expression space. Traceable distillation
must choose from a finite set of real cells and original genes. Sampling,
sketching, and medoid methods favor strata or local coverage. Their average
coverage may still leave a few biological directions poorly preserved. We
therefore use characteristic functions to select real cells. Fixed-CF is a
simple uniform weight baseline. Minmax-CF adds adaptive worst direction
weights. It uses the same discrete rule for cells and genes.

Minmax-CF is designed for annotated single-cell training collections. Such
annotations may identify cell type, condition, or another biological
stratum. They define strata and quotas so large groups do not determine the
whole coreset. Test groups and downstream outcomes do not enter selection.

Single-cell compression also has a second axis: genes. Removing cells changes
coverage of biological states, donors, technologies, conditions, and rare
populations. Removing genes changes marker support, pathway support,
sparsity, and model inputs. Frozen SCimilarity heads make this difference
clear. Removed genes are set to zero in a fixed vocabulary
\citep{heimberg2025scimilarity}. We therefore keep separate cell and gene
budgets. Total compression is evaluated with cell and gene allocations, not
with one ratio alone.

We evaluate the distilled data rather than only the selector objective.
Preprocessing and selection are fitted inside each training split. Under
matched cell, gene, and training budgets, we compare Minmax-CF with Full,
synthetic PCA-Centroid, and synthetic DM. Fixed-CF tests
adaptive weighting as a simple internal baseline. Anti-cell is an adverse
control for the cell score. The evaluation separates two questions. The
first asks how much downstream utility the compressed data retain. The
second asks whether the selected entries remain linked to measured evidence.

Our main contributions are:
\begin{itemize}
    \item \textbf{Utility-first Traceable Coresets:} we formulate dual-axis compression as worst-direction CF subset selection. Our contribution is not merely retaining the indices of a real subset. It lies in achieving competitive downstream utility while preserving traceability.

    \item \textbf{Discrete Min--Max Cell and Gene Selection:} We propose \textbf{Minmax-CF}. Its adaptive characteristic function weights emphasize poorly preserved directions. The method selects measured cells and genes.

    \item \textbf{Utility and Source Evaluation Across Datasets:} We evaluate matched compression budgets across datasets. The results report mean utility together with direct inspection through retained cell IDs
    and gene symbols.
\end{itemize}

%% file: sections/related_work.tex
\section{Related Work}

\paragraph{Dataset distillation outputs.}
Compression methods return synthetic examples, metacells, or observed
subsets. Synthetic methods optimize examples through gradients,
training paths, or feature distributions
\citep{wang2018distillation,zhao2021gradient,cazenavette2022trajectory,zhao2023distribution}.
Metacells aggregate measured neighborhoods, while observed subsets retain
source rows. NCFM learns CF directions to synthesize samples\citep{wang2025ncfm}. We instead fix the directions and select observed rows/columns under an entropy-regularized worst-direction weight. Every output row and column keeps its
source identifier. Synthetic PCA-Centroid and DM provide
matched controls for this output difference, which are not intended to rank every generator.

\paragraph{Single-cell sketching and gene selection.}
scDD is the closest single-cell distillation method \citep{yu2025scdd}. However, it emits synthetic diffusion profiles rather than observed rows. It is thus a generator, not a matched observed-subset control, so we position it in text. Foundation model features guide this process. A generated profile
does not retain the donor, count vector, or experimental context of one
assayed cell. Geometric sketching covers transcriptional state space
\citep{hie2019geometric}. SEACells aggregates measured neighborhoods into metacells \citep{persad2023seacells}. PCA-Centroid control averages partitions defined for each class. Gene selection changes feature support rather than cell support. Marker panels and detection
frequency filters are common examples. Gene selection can change sparse zero
patterns. It may also remove rare markers without changing cell identities.
We therefore analyze cell and gene budgets separately.

\paragraph{Single-cell foundation models and efficient adaptation.}
Single-cell foundation models provide reusable representations. A frozen
encoder with a linear head differs from encoder pretraining or fine tuning.
We study a fixed representation. The encoder computes embeddings once, and
we fit downstream heads to the stored values. Adapter methods and full
encoder tuning remain separate settings. Our results do not cover them.

\paragraph{Biological shifts.}
Shuffled cell splits can leak donors, platforms, or perturbation conditions
through preprocessing and selection. We fit every transformation on the
training split and hold out the target groups. Norman Perturb-seq
\citep{Norman2019} separates combinations by component visibility.
Additive predictors remain competitive for unseen components
\citep{ahlmanneltze2025linear}. Distribution fidelity, local geometry, and prediction of unseen perturbations answer different questions. We therefore report distribution fidelity and perturbation prediction separately.

%% file: sections/method.tex
\begin{figure*}[t]
\centering
\includegraphics[width=0.98\textwidth]{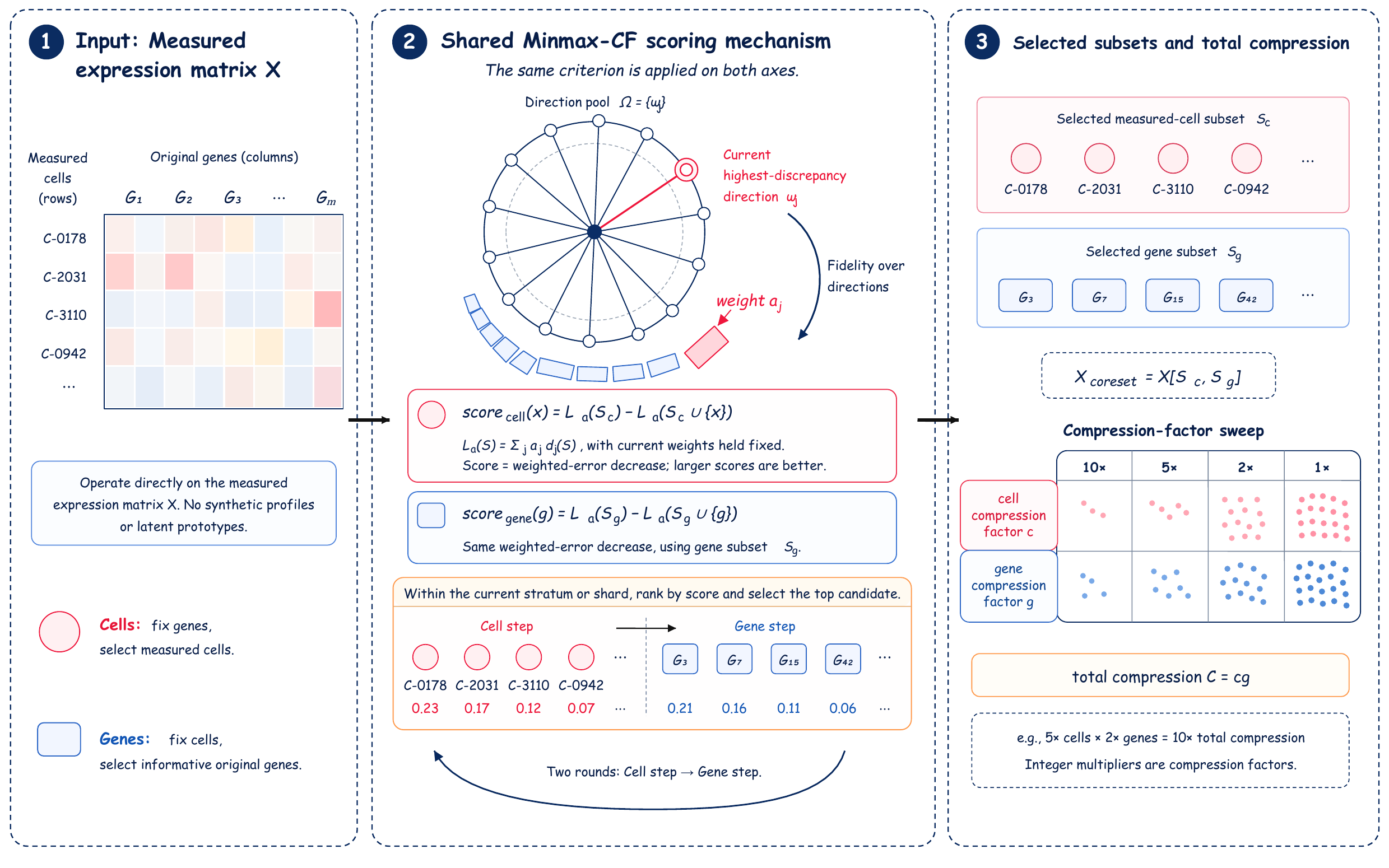}
\caption{\textbf{Minmax-CF scoring and selection.} At each greedy step,
candidates are ranked by the decrease in weighted CF error. We hold direction
weights fixed during comparison and apply the rule to both axes.}
\label{fig:method-schematic}
\end{figure*}
\section{Method}

\subsection{Problem Setup and Traceable Output}
Let \(G=(g_1,\ldots,g_p)\) be the ordered gene list and
\(X\in\mathbb{R}^{n\times p}\) the training expression matrix. Cell \(i\)
has expression profile \(x_i\) and biological stratum label \(c_i\). We
write the training data as
\[
D^{\mathrm{tr}}=\{(x_i,c_i)\}_{i=1}^{n}.
\]
The selector uses \(c_i\) to define strata and quotas. It therefore requires
labels. We do not evaluate selection without labels. Metadata does not enter
the score. It is retrieved through the source cell index. The cell and gene
budgets are \(k_c\) and \(k_g\). The selector returns
\(S\subseteq[n]\) and \(T\subseteq[p]\). Their sizes are \(k_c\) and
\(k_g\). Its output is
\[
\widetilde X=X[S,T],\qquad
C=\frac{np}{k_c k_g},
\]
together with the source cell indices and labels
\(\{(i,c_i):i\in S\}\), their metadata, and retained gene symbols
\(\{g_j:j\in T\}\). Every value remains linked to a source cell and
gene. The method releases no synthetic profile or latent prototype. The
factor \(C\) measures the reduction in dense matrix entries. It does not
measure file size, selection cost, or speed.

Each selected row and column keeps its source index and can be retrieved
directly.

We fix the train, validation, and test split before selection. Normalization
parameters come from the training data. The same rule applies to feature
representations, strata, SVD coordinates, gene scores, and selected indices.
Validation data may choose a listed budget. Test donors, technologies, and
perturbation groups remain untouched.

\subsection{CF Discrepancy for Cells and Genes}
Minmax-CF applies one CF selection rule to two discrete supports:
cells and genes. Let \(\mathcal V_s\) be the candidates in stratum \(s\).
Each candidate has a representation \(z(v)\in\mathbb{R}^d\) built from
training data. On the cell axis, \(\mathcal V_s=\{i:c_i=s\}\), and the
representation uses the current genes. On the gene axis, a gene is
represented by its expression across the selected cells. We draw a fixed and
seeded Gaussian direction pool
\(\Omega=\{\omega_j\}_{j=1}^{M}\), where
\(\omega_j\overset{\mathrm{iid}}{\sim}\mathcal N(0,I_d/d)\).
Here, \(I_d\) is the \(d\times d\) identity matrix. The factor \(1/d\)
sets the expected squared norm to one. For nonempty
\(A\subseteq\mathcal V_s\), define
\[
\begin{aligned}
\Phi_A(\omega_j)
&=\frac{1}{|A|}\sum_{v\in A}
\exp\!\left(\mathrm{i}\,\omega_j^\top z(v)\right),\\
d_j(Q)
&=\left|\Phi_Q(\omega_j)-\Phi_{\mathcal V_s}(\omega_j)\right|^2.
\end{aligned}
\]
The value \(d_j(Q)\) compares the empirical CF of the selected set with that
of its full stratum along direction \(j\). A smaller value means closer
agreement in that direction. The direction pool is fixed within each
comparison. Candidate rankings therefore use the same directions.
Fixed-CF uses greedy selection to approximately minimize
\begin{equation}
\min_{\substack{Q\subseteq\mathcal V_s\\|Q|=k}}
\mathcal L_{\mathrm{fix}}(Q),\qquad
\mathcal L_{\mathrm{fix}}(Q)=\frac{1}{M}\sum_{j=1}^{M}d_j(Q),
\label{eq:fixed-cf}
\end{equation}
which gives every direction equal weight.

\subsection{Entropy Regularization for Minmax-CF}
Uniform averaging may hide a few poorly matched projections. Minmax-CF
instead uses an entropy regularized worst direction
objective
\begin{equation}
\min_{\substack{Q\subseteq\mathcal V_s\\|Q|=k}}
\max_{a\in\Delta_M}
\sum_{j=1}^{M}a_jd_j(Q)-\tau\,\mathrm{KL}(a\|u),
\label{eq:minmax-cf}
\end{equation}
where \(\Delta_M=\{a\in\mathbb R_+^M:\sum_j a_j=1\}\), \(u_j=1/M\), and
\(\tau>0\). For fixed \(Q\), the inner maximization gives
\begin{equation}
\begin{aligned}
a_j(Q)
&=\frac{\exp(d_j(Q)/\tau)}
{\sum_{\ell=1}^{M}\exp(d_\ell(Q)/\tau)},\\
\mathcal L_\tau(Q)
&=\tau\log\!\left[
\frac{1}{M}\sum_{j=1}^{M}\exp(d_j(Q)/\tau)\right].
\end{aligned}
\label{eq:adversary}
\end{equation}
Directions with larger current discrepancies receive more weight. The KL
term limits concentration on one direction that may be noisy. As
\(\tau\to0\), \(\mathcal L_\tau\) approaches the largest directional
discrepancy. Larger values of \(\tau\) make the weights more even. Fixed-CF
keeps them exactly uniform.

Direct optimization of Equation~\ref{eq:minmax-cf} requires a combinatorial
search. We use a greedy approximation.
The initial set is \(Q_0=\varnothing\). We set \(a^{(0)}=u\) because the CF
of an empty set is undefined. At step \(t\), weights come from \(Q_t\). They
remain fixed while candidates are compared:
\begin{equation}
\begin{aligned}
v_{t+1}
&\in\arg\min_{v\in\mathcal C_s\setminus Q_t}
\sum_{j=1}^{M}a_j^{(t)}d_j(Q_t\cup\{v\}),\\
Q_{t+1}
&=Q_t\cup\{v_{t+1}\},
\end{aligned}
\label{eq:greedy-step}
\end{equation}
where \(a^{(t)}=a(Q_t)\) for \(t>0\) and \(\mathcal C_s\) is the
candidate pool from training data. Fixed-CF uses the same rule but sets
\(a^{(t)}=u\) at every step. The two methods therefore differ only in
whether the direction weights are updated.

For each candidate, the rule evaluates the weighted discrepancy after adding
that candidate. It selects the candidate with the smallest resulting value.
The direction weights are then updated from the enlarged set before the next
step. Holding weights fixed during one comparison gives every candidate the
same current objective.

\input{tables/cross_dataset_transfer}

\subsection{Alternating Cell and Gene Selection}
We apply the rule to both axes for two rounds. The first cell step uses all
eligible genes. In each round, cell representations are rebuilt from the
genes retained in the previous step. The selector then returns exactly
\(k_c\) measured cells. Gene representations are rebuilt from these cells,
and the selector returns exactly \(k_g\) original genes. The second round
repeats both steps with the updated subset from the other axis.

Alternation couples the two budgets. A new gene subset changes the cell
representation used in the next round. A new cell subset likewise changes
the expression profile used to compare genes. Two rounds apply this feedback
once in each direction while keeping the procedure fixed across datasets.

On both axes, we first normalize item vectors by their L2 norm. We project
them with an SVD fitted on training data. The SVD has at most 64 components.
We then normalize the projected vectors again. Cells use \(c_i\) strata in
lexical order. Genes are ordered by current mean expression and split into
at most eight equal array strata. Gene index resolves ties.

For stratum size \(n_s\), we initialize
\(k_s=\max\{1,\lfloor k n_s/\sum_r n_r\rfloor\}\) and clip it at \(n_s\).
Residual quotas follow decreasing fractional remainder. Stratum index
resolves ties. If excess quotas must be removed, we reverse this order.
A quota above 50 is permuted with the registered seed. It is then divided
into \(\lceil k_s/50\rceil\) shards by the same allocation rule.

The requested axis budget \(k\) must be at least the number of nonempty
strata. Every registered budget meets this condition. We solve
Equation~\ref{eq:minmax-cf} independently within each stratum. A sharded
stratum is solved within each shard. Shard outputs are united within a
stratum, and stratum outputs form the axis subset. We do not rank candidates
across strata.

We use \(M=512\) and \(\tau=0.05\). All seeded operations use the registered
seed for that run. Each shard has a candidate limit of 700. A limited pool
combines at most \(\min\{350,\max(3k_s,80)\}\) medoids with samples from the
same seed. An uncapped pool contains all items. Greedy ties follow source
order. The trace stores the axis, round, stratum, source index, and
discrepancy before and after selection. Labels define strata only. Test
outcomes and downstream losses never enter the score.

The candidate limit bounds the cost of each greedy step.

\subsection{Baselines and Budget Allocation}
We use Fixed-CF as a simple baseline proposed to isolate adaptive weighting.
It shares the support, strata, quotas, candidate pools, budgets, and greedy
rule with Minmax-CF. Only its direction weights remain uniform. Fixed-CF
therefore preserves source cell IDs and original gene symbols.

The external controls are synthetic PCA-Centroid and synthetic DM. They form
profiles for each class in PCA space or by matching
feature distributions. Neither maps a profile to one assayed cell. All
controls share training labels, strata, quotas, preprocessing, budgets, and
head training.

We report the retained cell count \(k_c\) and gene count \(k_g\) for every
experiment. The hPancreas study covers 20 cell and gene allocations from
\(16\times\) to \(256\times\) total compression. Synthetic controls match
both counts and the training budget. Cell compression changes population
and rare state coverage. Gene compression changes marker, pathway, and
sparsity support. The grid measures this allocation effect rather than
defining one general ratio. Frozen encoder experiments form a separate
setting. They select cells only and keep the full gene vocabulary.

%% file: tables/cross_dataset_transfer.tex
\begin{table*}[!t]
\centering

\setlength{\tabcolsep}{3.2pt}
\renewcommand{\arraystretch}{0.90}
\begin{tabular*}{\textwidth}{@{\extracolsep{\fill}}llccccc@{}}
\toprule[1.5pt]
\textbf{Total} & \textbf{Method} & \textbf{HP14} & \textbf{Kim Lung} &
\textbf{Zheng68K} & \textbf{Liver} & \textbf{Jerber iPSC} \\
\midrule
\rowcolor{gray!18}
\multicolumn{7}{c}{\textit{Full reference}} \\
$1\times$ & Full & 86.5 & 95.9 & 85.5 & 96.8 & 81.4 \\
\rowcolor{gray!12}
\multicolumn{7}{c}{\textit{Total compression $16\times$}} \\
$16\times$ & Synthetic PCA-Centroid & 85.3 & 88.5 & 66.1 & 74.2 & 76.6 \\
$16\times$ & DM & 70.1 & 83.2 & 59.1 & 92.3 & 54.5 \\
$16\times$ & \textbf{Minmax-CF} & \textbf{88.3} & \textbf{93.1} & \textbf{79.5} & \textbf{94.6} & \textbf{76.9} \\
\rowcolor{gray!12}
\multicolumn{7}{c}{\textit{Total compression $32\times$}} \\
$32\times$ & Synthetic PCA-Centroid & 84.1 & 87.6 & 66.0 & 73.4 & 76.2 \\
$32\times$ & DM & 77.4 & 83.4 & 59.1 & \textbf{93.0} & 53.8 \\
$32\times$ & \textbf{Minmax-CF} & \textbf{86.2} & \textbf{95.0} & \textbf{76.8} & 92.7 & \textbf{77.6} \\
\rowcolor{gray!12}
\multicolumn{7}{c}{\textit{Total compression $64\times$}} \\
$64\times$ & Synthetic PCA-Centroid & 82.8 & 87.7 & 64.9 & 73.4 & 75.5 \\
$64\times$ & DM & 77.3 & 84.2 & 59.5 & 92.7 & 54.3 \\
$64\times$ & \textbf{Minmax-CF} & \textbf{84.2} & \textbf{97.5} & \textbf{80.3} & \textbf{94.1} & \textbf{76.4} \\
\rowcolor{gray!12}
\multicolumn{7}{c}{\textit{Total compression $128\times$}} \\
$128\times$ & Synthetic PCA-Centroid & 78.3 & 84.2 & 64.8 & 78.6 & 74.6 \\
$128\times$ & DM & 65.2 & 83.1 & 58.9 & 90.3 & 54.9 \\
$128\times$ & \textbf{Minmax-CF} & \textbf{78.5} & \textbf{93.0} & \textbf{77.1} & \textbf{92.5} & \textbf{76.3} \\
\rowcolor{gray!12}
\multicolumn{7}{c}{\textit{Total compression $256\times$}} \\
$256\times$ & Synthetic PCA-Centroid & 79.7 & 83.0 & 68.7 & 77.8 & \textbf{74.2} \\
$256\times$ & DM & 76.7 & 81.7 & 60.4 & 90.8 & 55.1 \\
$256\times$ & \textbf{Minmax-CF} & \textbf{81.2} & \textbf{92.1} & \textbf{74.0} & \textbf{94.2} & 73.8 \\
\bottomrule[1.5pt]
\end{tabular*}
\caption{\textbf{Coarse lineage performance across datasets.}
Test macro-F1 is shown in percent as the mean across registered seeds. Per
seed values and variation are omitted for readability. Synthetic
PCA-Centroid and DM are controls built for each class. At each compression,
methods share labels, strata, preprocessing, budgets, and training. Bold
marks the best compressed mean.}
\label{tab:cross-dataset-transfer}
\end{table*}

%% file: sections/experiments.tex
\section{Experiments}

\subsection{Experimental Setup}

\paragraph{Datasets, splits, and methods.}
Table~\ref{tab:cross-dataset-transfer} evaluates five benchmarks. HP14 uses
a source OOD split with 14,813 cells and 3,000 genes. Kim
Lung~\citep{Kim2020Lung} uses a patient OOD split with 30,472 cells and
20,793 genes. Jerber iPSC~\citep{jerber2021population} uses a donor OOD split
with 205,416 cells and 32,738 genes. Zheng68K~\citep{Zheng2017} uses a
stratified IID split with 68,450 cells and 16,906 genes.
Liver~\citep{MacParland2018Liver} uses the same split type with 8,444 cells
and 20,007 genes.

Additional analyses use donor OOD MS~\citep{Schirmer2019}, technology OOD
hPancreas~\citep{Luecken2022}, Norman perturbations~\citep{Norman2019}, and
IFNB~\citep{Kang2018}. Preprocessing and selection use training data only.
The main external comparison uses Full, synthetic PCA-Centroid, synthetic
DM, and Minmax-CF. Fixed-CF is a simple internal baseline
for adaptive weighting. SCimilarity experiments select cells and genes, set
omitted genes to zero, and train only a linear head. All data were released
previously under their original access terms. We collected no new
participants or specimens. The original data access conditions still apply.

\paragraph{Metrics and evidence units.}
Classification reports balanced accuracy and macro-F1. Rare class recall
uses classes defined from the training split. Norman reports correlations
for gene and pathway effects. Each dataset in
Table~\ref{tab:cross-dataset-transfer} uses its registered split. The table reports arithmetic means across registered seeds. We report means over seeds. The reported gaps exceed one per-seed standard deviation, so ranking conclusions are stable.

Frozen linear heads use weighted cross entropy and AdamW. The learning rate
is \(5{\times}10^{-3}\), the weight decay is \(10^{-4}\), and the batch size
is 512. Training runs for at most 50 epochs. Early stopping begins after 8
epochs and uses patience 7 on validation macro-F1. The paired Minmax-CF and
Anti-cell analysis uses a full reference with 64 dimensions. It evaluates
1,024 CF directions on held out data and Reactome means after zero filling
\citep{milacic2024reactome}.

\subsection{Minmax-CF Across Cell and Gene Budgets}

We first examine how cell and gene allocation affects Minmax-CF under a
fixed total budget. Figure~\ref{fig:compression-scaling} reports 20 hPancreas
allocations from $16\times$ to $256\times$ compression. A larger $c/g$
retains fewer cells and more genes. Macro-F1 reaches 79.2\%, 79.2\%, and
82.2\% at $16\times$, $32\times$, and $64\times$.

The trend is not monotonic at every budget. The $128\times$ group peaks at
c16/g8. The $256\times$ group is also not monotonic, although c64/g4 is its
best setting. Gene retention often helps in this study, but the results do
not define a general ratio.

We next compare Minmax-CF with matched external controls across datasets.
Table~\ref{tab:cross-dataset-transfer} reports coarse lineage macro-F1.
Synthetic PCA-Centroid forms centroids for each class in PCA space.
Synthetic DM forms generated profiles for each class.
Neither maps a profile to one assayed cell. All methods share training
labels, strata, quotas, preprocessing, and cell, gene, and training budgets.

Minmax-CF exceeds synthetic PCA-Centroid and synthetic DM
in 24 of 25 comparisons against each control. Across the 25 displayed means,
its relative advantage is 10.4\% over synthetic PCA-Centroid and 17.4\% over
synthetic DM. The exceptions are $256\times$ Jerber and
$32\times$ Liver, respectively. These means do not show a gain in every
setting.

\subsection{Compression Scaling and Allocation Sensitivity}

\begin{figure}[t]
\centering
\includegraphics[width=0.96\columnwidth]{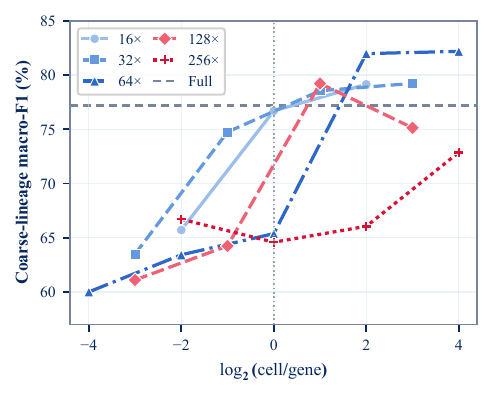}
\caption{\textbf{Cell and gene allocation performance.} Minmax-CF test
macro-F1 for 20 allocations. Here, \(c/g\) divides cell compression by gene
compression. Lines join measured settings within each total. The dashed line
is Full.}
\label{fig:compression-scaling}

\end{figure}

Figure~\ref{fig:compression-scaling} connects observed allocations within
each total. Macro-F1 spans 63.5\% to 79.2\% at $32\times$, 60.0\% to 82.2\%
at $64\times$, and 61.1\% to 79.2\% at $128\times$. Several allocations
meet or exceed the 77.2\% Full reference. All $256\times$ settings remain
below it. These ranges show that performance depends on how the budget is
divided across cells and genes.

Allocation clearly affects performance. Passing Full may reflect
regularization or limited training. It does not mean that the compressed
matrix contains more information. The curves change direction, so one cell to
gene ratio does not fit every budget.

\subsection{Minmax-CF Selection}

Minmax-CF uses the same adaptive CF score on measured cells and original
genes. Fixed-CF is a simple baseline that keeps direction weights uniform.
The matched $64\times$ analysis compares these rules.
Minmax-CF reduces soft CF error on test directions by 42.2\% on hPancreas,
47.6\% on MS, and 48.0\% on Norman. The reductions in worst direction error
are 44.3\%, 50.0\%, and 51.7\%. The comparison uses 27 fold and seed units
for hPancreas and 3 units for each other dataset. This comparison isolates
the effect of adaptive weights on the CF criterion. It does not establish a gain on every downstream task. Because the candidate pools, budgets, representations, and greedy updates are matched, these reductions isolate the contribution of adaptive direction weighting. Fixed-CF is therefore the uniform-weight variant of our framework, whereas the subsequent experiment tests whether the candidate ordering induced by the adaptive score is informative.

Table~\ref{tab:anti-cell-ablation} gives a separate reversal control.
Anti-cell uses the Minmax-CF adaptive score but deliberately selects the
cell candidate with the lowest score. This candidate is expected to reduce
the weighted CF error least. The gene rule, labels, strata, quotas, budgets,
and downstream training remain fixed. Minmax-CF has higher macro-F1 on all
seven classification tasks. This result shows that the order induced by the
cell score contains information useful for these tasks. It also makes shared
budgets and class quotas less plausible as the sole explanation. As an
adverse control, Anti-cell does not rule out every other cause.

Relative to Anti-cell, Minmax-CF reduces soft CF error on test directions by
18.6\%, 32.9\%, and 21.0\%. It reduces worst direction error by 22.4\%,
21.7\%, and 36.3\%. Reactome pathway MAE
\citep{milacic2024reactome} improves on hPancreas and MS but is 2.3\% higher
on Norman. The pathway result is therefore mixed.

\input{tables/anti_cell_ablation}

\begin{figure}[!b]
\centering
\includegraphics[width=\columnwidth]{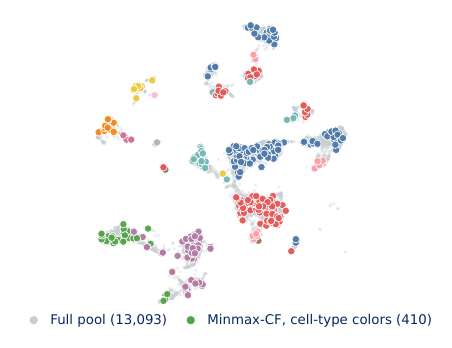}
\caption{\textbf{Selected cell manifold coverage at $64\times$.} The full
training pool is gray, with 410 selected source cells overlaid. This is a
qualitative coverage view, not source attribution.}
\label{fig:hpancreas-manifold}
\end{figure}

\subsection{Biological Traceability Analysis}

\paragraph{Cell manifold coverage.}
Minmax-CF rows link to the original annotations and SCimilarity embeddings
\citep{heimberg2025scimilarity}. Figure~\ref{fig:hpancreas-manifold} overlays
the 410 cells retained by the $64\times$ c32/g2 allocation on the training
pool. They occupy its major regions, and their source locations can be
inspected in the common embedding. The selector does not use UMAP
coordinates. This panel is qualitative and does not prove coverage of every
rare state.

\begin{figure}[!t]
\centering
\includegraphics[width=\columnwidth]{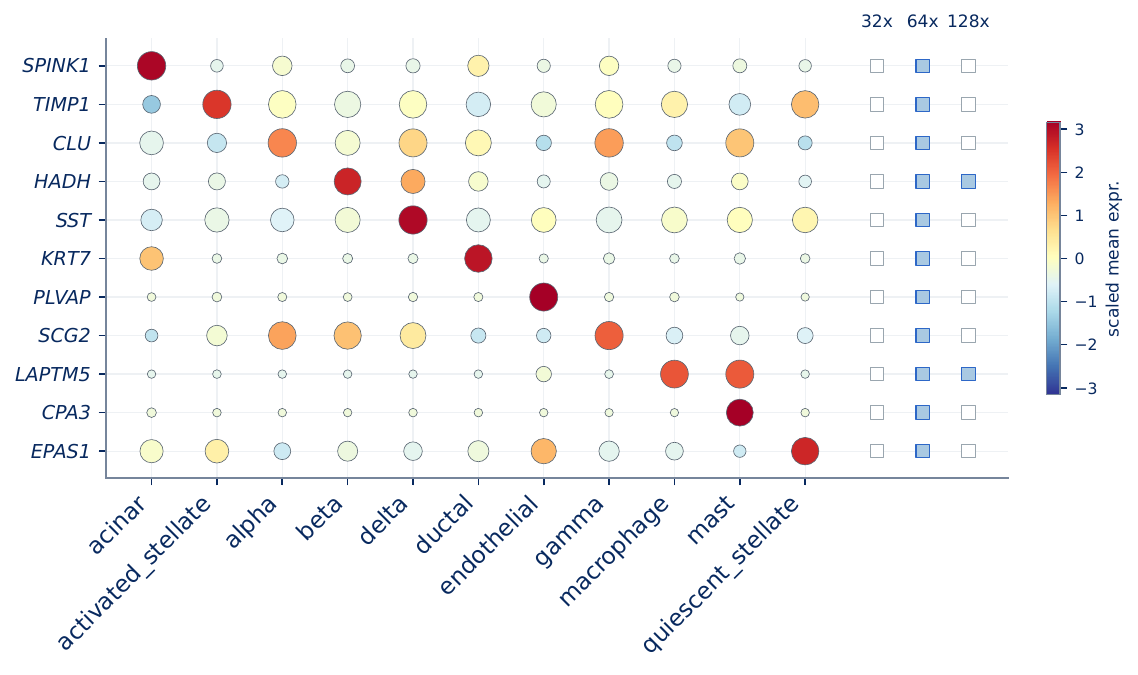}
\caption{\textbf{Illustrative retained gene lookup.} Displayed genes were
chosen from the $64\times$ set, so its column is filled. Dot area shows the
expressing cell fraction, and color shows scaled mean expression. Filled squares mark retention, while open squares mark exclusion. Only HADH and LAPTM5 remain at $128\times$.}
\label{fig:selected-gene-markers}
\end{figure}

\paragraph{Selected gene interpretability.}
Figure~\ref{fig:selected-gene-markers} displays retained gene symbols. It
shows one retained gene for every cell type with at least 30 training cells.
The genes were chosen after $64\times$ selection by one versus rest
specificity.

The set spans exocrine, endocrine, vascular, stromal, and immune programs.
Examples include SPINK1, SST, PLVAP, and CPA3. The panel shows that retained
symbols can be inspected. It does not show better marker or pathway
preservation. The displayed genes were chosen only after selection. Gene
identities are not nested across budgets, and zero filling does not support a
pathway fidelity claim.

\paragraph{Perturbation performance.}
On Norman, Minmax-CF exceeds synthetic PCA-Centroid and Fixed-CF but remains below Full. This result complements the held-out CF analysis by testing whether preservation of observed distribution directions transfers to perturbation
responses. The compressed methods recover part of the response structure, while Full retains the strongest gene-level and pathway-level correlations.
Thus, the Norman experiment connects the direct Minmax-CF objective with a distinct downstream task.

\paragraph{Allocations rich in genes under frozen encoders.}
SCimilarity \citep{heimberg2025scimilarity} uses a fixed gene vocabulary.
Removed genes enter the frozen encoder as zeros. At a fixed budget, a larger
$c/g$ retains fewer cells and more genes. Figure~\ref{fig:compression-scaling}
shows higher macro-F1 at moderate compression. Stronger compression can instead
limit cell coverage, so the two budgets remain separate.

\input{tables/hpancreas_budget_semantics}

\subsection{Evidence Boundaries}

\begin{figure}[!t]
\centering
\includegraphics[width=0.95\columnwidth]{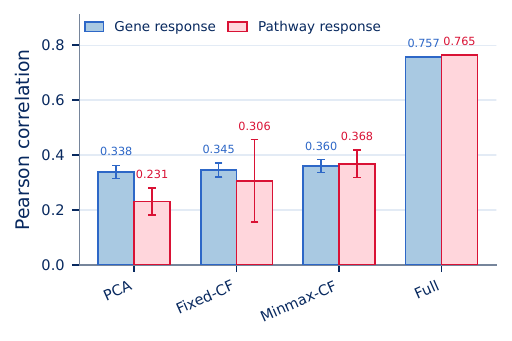}
\caption{\textbf{Norman perturbation performance.} Pearson correlations between predicted and observed test deltas. Bars show means and sample standard deviations over three seeds. PCA denotes synthetic PCA-Centroid.}

\label{fig:perturbation-recovery}
\end{figure}

\paragraph{Scope of the evidence.}
Table~\ref{tab:anti-cell-ablation} links the cell score to matched downstream
readouts. Figures~\ref{fig:compression-scaling} and
\ref{fig:perturbation-recovery} show that allocation and task affect the
remaining utility. The results support separate cell and gene budgets. They
also support subset selection that preserves source links. Fixed-CF isolates
adaptive weighting only for CF error on test directions.
Table~\ref{tab:cross-dataset-transfer} reports means without variation, so it
does not show stable gains in every run. Minmax-CF also requires training
labels for strata and quotas. It does not guarantee prediction of
perturbations absent from training. With a fixed gene vocabulary, zero
filling can make budget allocation more important than the selector.
\paragraph{Traceability enables biological inspection.}
Minmax-CF outputs measured cell IDs and original gene symbols rather than
latent profiles. This makes the distilled set auditable beyond aggregate
accuracy: Figure~\ref{fig:hpancreas-manifold} inspects source-cell coverage in
an independent embedding, while Figure~\ref{fig:selected-gene-markers} checks
whether retained genes preserve interpretable marker support. The key contribution is therefore not index retention alone, while maintaining competitive utility while keeping the selected evidence biologically traceable.

%% file: tables/anti_cell_ablation.tex
\begin{table*}[!t]
\centering
\setlength{\tabcolsep}{6.0pt}
\renewcommand{\arraystretch}{0.96}
\begin{tabular}{lccc|ccc}
\toprule[1.5pt]
\textbf{Dataset / task} &
\multicolumn{3}{c|}{\textbf{Fidelity error: Minmax-CF / Anti-cell}} &
\multicolumn{3}{c}{\textbf{Macro-F1 in percent}} \\
\cmidrule(lr){2-4} \cmidrule(lr){5-7}
& \textbf{Soft CF $\downarrow$} &
\textbf{Worst direction CF $\downarrow$} &
\textbf{Pathway MAE $\downarrow$} &
\textbf{Minmax-CF $\uparrow$} &
\textbf{Anti-cell $\uparrow$} &
\textbf{Gain} \\
\midrule
hPancreas / cell type
& \textbf{0.0032} / 0.0039 & \textbf{0.0293} / 0.0377 & \textbf{0.2535} / 0.2545
& \textbf{44.3} & 36.3 & $+8.0$ \\
hPancreas / coarse lineage
& \textbf{0.0032} / 0.0039 & \textbf{0.0293} / 0.0377 & \textbf{0.2535} / 0.2545
& \textbf{69.3} & 64.2 & $+5.2$ \\
MS / cell type
& \textbf{0.0078} / 0.0116 & \textbf{0.0831} / 0.1061 & \textbf{0.4393} / 0.5189
& \textbf{30.2} & 29.0 & $+1.3$ \\
MS / condition
& \textbf{0.0078} / 0.0116 & \textbf{0.0831} / 0.1061 & \textbf{0.4393} / 0.5189
& \textbf{61.8} & 49.4 & $+12.5$ \\
Norman / fidelity
& \textbf{0.0118} / 0.0149 & \textbf{0.1145} / 0.1798 & 0.5126 / \textbf{0.5013}
& -- & -- & -- \\
IFNB / cell type
& -- & -- & -- & \textbf{57.4} & 48.8 & $+8.6$ \\
IFNB / stimulation
& -- & -- & -- & \textbf{89.7} & 85.4 & $+4.3$ \\
IFNB / combined label
& -- & -- & -- & \textbf{48.7} & 37.8 & $+10.9$ \\
\bottomrule[1.5pt]
\end{tabular}
\caption{\textbf{Cell score reversal control at $64\times$ c8/g8.}
Anti-cell uses the same adaptive cell score but selects the candidate with
the lowest score. Gene selection and budgets stay fixed. Fidelity repeats for tasks
that share a subset. Gains use unrounded macro-F1. Bold marks the better
value.}
\label{tab:anti-cell-ablation}
\end{table*}

%% file: tables/hpancreas_budget_semantics.tex
\paragraph{Budget semantics.}
The $16\times$ allocation retains 6,547 cells and 2,387 genes. At
$32\times$, it retains 1,637 cells and 4,774 genes. The $64\times$
allocation retains 410 cells and 9,547 genes. At $128\times$, it retains
819 cells and 2,387 genes.
At $256\times$, it retains 410 cells and 2,387 genes. Total compression
multiplies the two axis factors. It measures dense matrix entries, not file
size or speed. Counts are not monotonic because these are distinct
allocations.

%% file: sections/conclusion.tex
\section{Conclusion}
We presented \textbf{Minmax-CF}, a traceable single-cell data distillation method for annotated training collections. It selects measured cells and original
genes through discrete min--max optimization, applying the same adaptive worst-direction CF rule to both axes so that selection concentrates on directions least well retained by the current coreset. Every retained entry stays linked to the source matrix and metadata. Matched synthetic PCA-Centroid and Distribution Matching (DM) controls show that generated summaries can be compact and useful, yet they do not preserve a one-to-one link to an assayed cell. Minmax-CF keeps this link while attaining higher mean macro-F1 in most tested settings. The allocation study shows that
cell and gene budgets require separate choices, and the Anti-cell control shows that the cell ranking affects downstream results. Our code will be made public.

%% file: references.bib
@misc{wang2018distillation,
  title = {Dataset Distillation},
  author = {Wang, Tongzhou and Zhu, Jun-Yan and Torralba, Antonio and Efros, Alexei A.},
  year = {2018},
  eprint = {1811.10959},
  archivePrefix = {arXiv},
  doi = {10.48550/arXiv.1811.10959}
}

@inproceedings{zhao2021gradient,
  title = {Dataset Condensation with Gradient Matching},
  author = {Zhao, Bo and Mopuri, Konda Reddy and Bilen, Hakan},
  booktitle = {International Conference on Learning Representations},
  year = {2021}
}

@inproceedings{cazenavette2022trajectory,
  title = {Dataset Distillation by Matching Training Trajectories},
  author = {Cazenavette, George and Wang, Tongzhou and Torralba, Antonio and Efros, Alexei A. and Zhu, Jun-Yan},
  booktitle = {Proceedings of the IEEE/CVF Conference on Computer Vision and Pattern Recognition},
  pages = {10718--10727},
  year = {2022},
  doi = {10.1109/CVPR52688.2022.01045}
}

@article{jerber2021population,
  title   = {Population-scale single-cell {RNA-seq} profiling across dopaminergic neuron differentiation},
  author  = {Jerber, Julie and Seaton, Daniel D. and Cuomo, Anna S. E. and others},
  journal = {Nature Genetics},
  volume  = {53},
  number  = {3},
  pages   = {304--312},
  year    = {2021},
  doi     = {10.1038/s41588-021-00801-6},
  url     = {https://doi.org/10.1038/s41588-021-00801-6}
}

@inproceedings{zhao2023distribution,
  title = {Dataset Condensation with Distribution Matching},
  author = {Zhao, Bo and Bilen, Hakan},
  booktitle = {Proceedings of the IEEE/CVF Winter Conference on Applications of Computer Vision},
  pages = {6514--6523},
  year = {2023}
}

@inproceedings{wang2025ncfm,
  title = {Dataset Distillation with Neural Characteristic Function: A Minmax Perspective},
  author = {Wang, Shaobo and Yang, Yicun and Liu, Zhiyuan and Sun, Chenghao and Hu, Xuming and He, Conghui and Zhang, Linfeng},
  booktitle = {Proceedings of the IEEE/CVF Conference on Computer Vision and Pattern Recognition},
  pages = {25570--25580},
  year = {2025}
}

@article{yu2025scdd,
  title = {{scDD}: {scRNA-seq} Dataset Distillation in Latent Codes with Single-Step Conditional Diffusion Generator},
  author = {Yu, Zhen and Han, Jianan and Liu, Yang and Chen, Qingchao},
  journal = {Knowledge-Based Systems},
  volume = {330},
  pages = {114610},
  year = {2025},
  doi = {10.1016/j.knosys.2025.114610}
}

@article{heimberg2025scimilarity,
  title = {A Cell Atlas Foundation Model for Scalable Search of Similar Human Cells},
  author = {Heimberg, Graham and Kuo, Tony and DePianto, Daryle J. and Salem, Omar and Heigl, Tobias and Diamant, Nathaniel and Scalia, Gabriele and Biancalani, Tommaso and Turley, Shannon J. and Rock, Jason R. and Corrada Bravo, H{\'e}ctor and Kaminker, Josh and Vander Heiden, Jason A. and Regev, Aviv},
  journal = {Nature},
  volume = {638},
  number = {8052},
  pages = {1085--1094},
  year = {2025},
  doi = {10.1038/s41586-024-08411-y}
}

@article{hie2019geometric,
  title = {Geometric Sketching Compactly Summarizes the Single-Cell Transcriptomic Landscape},
  author = {Hie, Brian and Cho, Hyunghoon and DeMeo, Benjamin and Bryson, Bryan and Berger, Bonnie},
  journal = {Cell Systems},
  volume = {8},
  number = {6},
  pages = {483--493.e7},
  year = {2019},
  doi = {10.1016/j.cels.2019.05.003}
}

@article{persad2023seacells,
  title = {{SEACells} Infers Transcriptional and Epigenomic Cellular States from Single-Cell Genomics Data},
  author = {Persad, Sitara and others},
  journal = {Nature Biotechnology},
  volume = {41},
  pages = {1746--1757},
  year = {2023},
  doi = {10.1038/s41587-023-01716-9}
}

@article{ahlmanneltze2025linear,
  title = {Deep-Learning-Based Gene Perturbation Effect Prediction Does Not Yet Outperform Simple Linear Baselines},
  author = {Ahlmann-Eltze, Constantin and Huber, Wolfgang and Anders, Simon},
  journal = {Nature Methods},
  year = {2025},
  doi = {10.1038/s41592-025-02772-6}
}

@article{milacic2024reactome,
  title = {The Reactome Pathway Knowledgebase 2024},
  author = {Milacic, Marija and others},
  journal = {Nucleic Acids Research},
  volume = {52},
  number = {D1},
  pages = {D672--D678},
  year = {2024},
  doi = {10.1093/nar/gkad1025}
}

@article{vandermaaten2008tsne,
  title = {Visualizing Data Using {t-SNE}},
  author = {van der Maaten, Laurens and Hinton, Geoffrey},
  journal = {Journal of Machine Learning Research},
  volume = {9},
  number = {86},
  pages = {2579--2605},
  year = {2008},
  url = {https://www.jmlr.org/papers/v9/vandermaaten08a.html}
}

@misc{mcinnes2018umap,
  title = {{UMAP}: Uniform Manifold Approximation and Projection for Dimension Reduction},
  author = {McInnes, Leland and Healy, John and Melville, James},
  year = {2018},
  eprint = {1802.03426},
  archivePrefix = {arXiv},
  doi = {10.48550/arXiv.1802.03426}
}

@article{Schirmer2019,
  author  = {Schirmer, Lucas and Velmeshev, Dmitry and
             Holmqvist, Staffan and Kaufmann, Max and
             Werneburg, Sebastian and Jung, Diane and
             Vistnes, Stephanie and Stockley, John H. and
             Young, Adam and Steindel, Maike and Tung, Brian and
             Goyal, Nitasha and Bhaduri, Aparna and Mayer, Simone and
             Engler, Jan Broder and Bayraktar, Omer A. and
             Franklin, Robin J. M. and Haeussler, Maximilian and
             Reynolds, Richard and Schafer, Dorothy P. and
             Friese, Manuel A. and Shiow, Lawrence R. and
             Kriegstein, Arnold R. and Rowitch, David H.},
  title   = {Neuronal vulnerability and multilineage diversity
             in multiple sclerosis},
  journal = {Nature},
  year    = {2019},
  volume  = {573},
  number  = {7772},
  pages   = {75--82},
  publisher = {Nature Publishing Group},
  doi     = {10.1038/s41586-019-1404-z}
}

@article{Luecken2022,
  author  = {Luecken, Malte D. and B{\"u}ttner, Maren and
             Chaichoompu, Kridsadakorn and Danese, Anna and
             Interlandi, Marta and Mueller, Michaela F. and
             Strobl, Daniel C. and Zappia, Luke and Dugas, Martin and
             Colom{\'e}-Tatch{\'e}, Maria and Theis, Fabian J.},
  title   = {Benchmarking atlas-level data integration in
             single-cell genomics},
  journal = {Nature Methods},
  year    = {2022},
  volume  = {19},
  number  = {1},
  pages   = {41--50},
  publisher = {Nature Publishing Group},
  doi     = {10.1038/s41592-021-01336-8}
}

@article{Kang2018,
  author  = {Kang, Hyun Min and Subramaniam, Meena and Targ, Sasha and
             Nguyen, Michelle and Maliskova, Lenka and
             McCarthy, Elizabeth and Wan, Eunice and Wong, Simon and
             Byrnes, Lauren and Lanata, Cristina M. and
             Gate, Rachel E. and Mostafavi, Sara and Marson, Alexander and
             Zaitlen, Noah and Criswell, Lindsey A. and Ye, Chun Jimmie},
  title   = {Multiplexed droplet single-cell {RNA}-sequencing using
             natural genetic variation},
  journal = {Nature Biotechnology},
  year    = {2018},
  volume  = {36},
  number  = {1},
  pages   = {89--94},
  publisher = {Nature Publishing Group},
  doi     = {10.1038/nbt.4042}
}

@article{Zheng2017,
  author  = {Zheng, Grace X. Y. and Terry, Jessica M. and Belgrader, Phillip and Ryvkin, Paul and Bent, Zachary W. and Wilson, Ryan and Ziraldo, Solongo B. and Wheeler, Tobias D. and McDermott, Geoff P. and Zhu, Junjie and Gregory, Mark T. and Shuga, Joe and Montesclaros, Luz and Underwood, Jason G. and Masquelier, Donald A. and Nishimura, Stefanie Y. and Schnall-Levin, Michael and Wyatt, Paul W. and Hindson, Christopher M. and Bharadwaj, Rajiv and Wong, Alexander and Ness, Kevin D. and Beppu, Lan W. and Deeg, H. Joachim and McFarland, Christopher and Loeb, Keith R. and Valente, William J. and Ericson, Nolan G. and Stevens, Emily A. and Radich, Jerald P. and Mikkelsen, Tarjei S. and Hindson, Benjamin J. and Bielas, Jason H.},
  title   = {Massively parallel digital transcriptional profiling of single cells},
  journal = {Nature Communications},
  year    = {2017},
  volume  = {8},
  number  = {1},
  pages   = {14049},
  publisher = {Nature Publishing Group},
  doi     = {10.1038/ncomms14049}
}

@article{MacParland2018Liver,
  author  = {MacParland, Sonya A. and Liu, Jeff C. and Ma, Xue-Zhong and Innes, Brendan T. and Bartczak, Agata M. and Gage, Blair K. and Manuel, Justin and Khuu, Nicholas and Echeverri, Juan and Linares, Ivan and others},
  title   = {Single cell {RNA} sequencing of human liver reveals distinct intrahepatic macrophage populations},
  journal = {Nature Communications},
  year    = {2018},
  volume  = {9},
  number  = {1},
  pages   = {4383},
  publisher = {Nature Publishing Group},
  doi     = {10.1038/s41467-018-06318-7}
}

@article{Norman2019,
  author  = {Norman, Thomas M. and Horlbeck, Max A. and Replogle, Joseph M. and Ge, Alex Y. and Xu, Albert and Jost, Marco and Gilbert, Luke A. and Weissman, Jonathan S.},
  title   = {Exploring genetic interaction manifolds constructed from rich single-cell phenotypes},
  journal = {Science},
  year    = {2019},
  volume  = {365},
  number  = {6455},
  pages   = {786--793},
  publisher = {American Association for the Advancement of Science},
  doi     = {10.1126/science.aax4438}
}

@article{Kim2020Lung,
  author  = {Kim, Nayoung and Kim, Hong Kwan and Lee, Kyungjong and Hong, Yourae and Cho, Jhingook H. and Choi, Jung Won and Lee, Jung-Il and Suh, Yeon-Lim and Ku, Bo Mi and Eum, Hye Hyeon and Choi, Soyean and Choi, Yoon-La and Joung, Je-Gun and Park, Woong-Yang and Jung, Hye Ryun and Sun, Jong-Mu and Lee, Se-Hoon and Ahn, Jin Seok and Park, Keunchil and Ahn, Myung-Ju and Lee, Hae-Ock},
  title   = {Single-cell {RNA} sequencing demonstrates the molecular and cellular reprogramming of metastatic lung adenocarcinoma},
  journal = {Nature Communications},
  year    = {2020},
  volume  = {11},
  number  = {1},
  pages   = {2285},
  publisher = {Nature Publishing Group},
  doi     = {10.1038/s41467-020-16164-1}
}
